\documentclass[12pt,english,reqno]{smfart}
\usepackage{mathptmx}
\usepackage{epsfig}
\theoremstyle{plain}

\theoremstyle{definition}

\renewcommand{\theequation}{\arabic{section}.\arabic{equation}}
\begin{document}
\title[CPA for disordered bosons]
{Coherent potential approximation for disordered bosons}
\author{S.E. Schmittner and M.R. Zirnbauer}
\date{December 23, 2010}
\begin{abstract}
A family of random models for bosonic quasi-particle excitations, e.g.\ the vibrations of a disordered solid, is introduced. The generator of the linearized phase space dynamics of these models is the sum of a deterministic and a random part. The former may describe any model of $N$ identical phonon bands, while the latter is a $d$-dimensional generalization of the random matrix model of Lueck, Sommers, and Zirnbauer (LSZ). The models are constructed so as to exclude the unphysical occurrence of runaway solutions. By using the Efetov-Wegner supersymmetry method in combination with the new technique of superbosonization, the disordered boson model is cast in the form of a supermatrix field theory. A self-consistent approximation of mean-field type arises from treating the field theory as a variational problem. The resulting scheme, referred to as a coherent potential approximation, becomes exact for large values of $N$. In the random-matrix limit, agreement with the results of LSZ is found. The self-consistency equation for the full $d$-dimensional problem is solved numerically.
%
\end{abstract}
\maketitle

\section{Introduction}

Small oscillations about the stable equilibrium of a many-body ground state are quantized as bosonic quasi-particles or bosons. In various physical contexts the linearized equations of motion for such excitations are known as the random phase approximation, or RPA equations for short
\cite{RingSchuck}. Concrete examples are furnished by the vibrational excitations of a solid, the spin waves of a magnet, the electromagnetic modes of an optical medium, or the density oscillations of a Bose-Einstein condensate.

Constrained by the requirement of dynamical stability, the Hamiltonian $H$ of any vibrational or quasi-boson system of the mentioned kind must lie in a positive cone, $\mathcal{E}$, of so-called elliptic symplectic generators. It should be stressed that although $H$ is Hermitian as an operator in Fock space, the quantum-to-classical mapping sends $H$ to an RPA generator $X$ which is in general neither Hermitian nor anti-Hermitian as a linear operator on the classical phase space. In view of this, a distinctive feature of the set of elliptic generators $X \in \mathcal{E}$ is that they can be brought to diagonal form (with real frequencies, corresponding to stable oscillatory motion) by real Bogoliubov transformations, i.e., by conjugating with elements of the real symplectic group $\mathrm{Sp}_\mathbb{R}\,$.

In this general setting, our goal is to investigate what happens with observables such as the spectral statistics and the transport properties when the bosonic system is strongly disordered. In particular, we wish to understand whether there exists some low-energy universality, possibly of an unusual type, due to the interplay between disorder and the geometry of the cone $\mathcal{E}$. (For example, a high degree of low-temperature universality is known to be observed \cite{Pohl} in strongly disordered solids as well as amorphous or glassy systems.) Motivated by this question, in the present paper we initiate the study of a class of semi-realistic random matrix models for disordered bosons.

By construction, the probability measures of the models we propose are supported on $\mathcal{E}$. Thus, unlike \cite{BBW09}, the unphysical behavior of runaway motion associated with complex frequencies is excluded. Our RPA generators $X = K + R$ have the particular feature of being sums of a deterministic and a random part. For simplicity we focus here on models without time-reversal symmetry, although TR-invariant models of a similar kind can be treated with little extra effort. By using a variant of the Efetov-Wegner supersymmetry method, we will derive an approximation for the density of states of mean-field or self-consistent type, reminiscent of the so-called `coherent potential approximation' (CPA) \cite{CPA-Korringa,CPA-Beeby,CPA-Soven}.

In the longer term, the goal is to develop a description of our disordered boson models by field theories of the non-linear sigma model type. (As is well known \cite{EM-RMP}, such a description has proven very useful for the case of disordered fermions). In that formulation, universality (if any) is expected to emerge whenever the renormalization group flow gets attracted to a few-parameter manifold of renormalizable field theories.

The plan of the paper is this. In Section \ref{sect:2} we outline the basic setting and in particular, we review the notion of positive cone of elliptic symplectic generators. We also introduce the random models to be considered and give a summary of the analytical results obtained.
Section \ref{sect:3} is concerned with the derivation of the coherent potential approximation for our models. For pedagogical reasons, we first discuss the zero-dimensional case in some detail. The extension to $d$ dimensions is given in Section \ref{sect:4}. There we also show some results for the numerical solution of the CPA equation.

\section{Setting, model, and results}\label{sect:2}
\setcounter{equation}{0}

In this section, we start with some background on mathematical formulation and describe a class of random models which are tractable by the superbosonization variant of the Efetov-Wegner supersymmetry method. We then give a summary of our analytical results, relegating the presentation of numerical results to the end of the paper.

\subsection{Setting}

Let $\{a_k^\dagger, a_k^{ \vphantom{\dagger}}\}_{k \in \Gamma}$ be a set of creation and annihilation operators for bosonic quasi-particles with quantum numbers $k \in \Gamma$. (For example, $\Gamma$ might be a discrete set of momenta selected by periodic boundary conditions in a finite box.) Such operators span a Hermitian symplectic vector space $W := \mathrm{span}_\mathbb{C} \{ a_k^{\vphantom{ \dagger}} , a_k^\dagger \}_{k \in \Gamma}$ with symplectic form $[ \, , \,] :\; W \times W \to \mathbb{C}$ defined by the canonical commutation relations
\begin{displaymath}
    [ a_k^{\vphantom{\dagger}} , a_{k^\prime}^\dagger ] =
    \delta_{k k^\prime} \;, \qquad [a_k , a_{k^\prime}] = 0 \,,
    \qquad [a_k^\dagger , a_{k^\prime}^\dagger] = 0 \,.
\end{displaymath}

We now assume that we are given a linear Hamiltonian dynamics on $W$. This may be interpreted either as a linear Hamiltonian flow on $W$ viewed as a classical phase space, or as a quantum time evolution on $W$ viewed as a subspace of the associative algebra of polynomials in $a_k^{ \vphantom{ \dagger}} $, $a_k^\dagger$ (the so-called Weyl algebra). In either case, the dynamical equations are
\begin{equation}\label{eq:RPA}
    \frac{d}{dt} \, a_k^\dagger =  \sum_{k^\prime} \big(a_{k^\prime}^\dagger Y_{k^\prime k} + a_{k^\prime} Z_{k^\prime k} \big) , \quad \frac{d}{dt}\, a_k =  \sum_{k^\prime} \big(a_{k^\prime}^\dagger \bar{Z}_{k^\prime k} + a_{k^\prime} \bar{Y}_{k^\prime k}\big) .
\end{equation}
(The bar means complex conjugation.) In order for the canonical commutation relations to be invariant under the dynamics, we require that $Y_{k k^\prime} = - \bar{Y}_{k^\prime k}$ and $Z_{k k^\prime} = Z_{k^\prime k}$. Thus $Y_{k k^\prime}$ are the matrix elements of an anti-Hermitian matrix $Y = - Y^\dagger$, while $Z_{k k^\prime}$ are those of a complex symmetric matrix $Z = Z^{\,\mathrm{t}}$. Altogether, these conditions mean that
\begin{equation}
    X := \begin{pmatrix} Y &\bar{Z} \cr Z &\bar{Y} \end{pmatrix}
\end{equation}
is the generator of a symplectic transformation. More precisely, defining the Lie algebra, $\mathfrak{sp}$, of the complex symplectic group by the linear condition
\begin{equation}\label{eq:def-sp-C}
    X = - J X^\mathrm{t} J^{-1} \;, \qquad J =
    \begin{pmatrix} 0 &\mathbf{1}\cr - \mathbf{1} &0 \end{pmatrix} ,
\end{equation}
$X$ lies in a non-compact real form $\mathfrak{sp}_\mathbb{R} \subset \mathfrak{sp}$ determined by
\begin{equation}
    X = - \Sigma_3 X^\dagger \Sigma_3 \;, \qquad \Sigma_3 = \begin{pmatrix} \mathbf{1} &0 \cr 0 &-\mathbf{1} \end{pmatrix} .
\end{equation}
It should be mentioned that this description is appropriate in the absence of time-reversal invariance. If time reversal is a symmetry of the physical system, then the time-evolution generator $X$ is subject to additional complex anti-linear conditions.

Equations (\ref{eq:RPA}) arise as the equations of motion for a system of non-interacting bosons with Hamiltonian
\begin{equation}\label{eq:2ndQ}
    H = \mathrm{i}\hbar \sum\nolimits_{k,\,k^\prime} \big( Y_{k k^\prime} a_k^\dagger a_{k^\prime}^{\vphantom{\dagger}} + {\textstyle{\frac{1} {2}}} Z_{k k^\prime} a_k a_{k^\prime} - {\textstyle{\frac{1}{2}}} \bar{Z}_{k k^\prime} a_k^\dagger a_{k^\prime}^\dagger \big)
\end{equation}
and dynamics $\mathrm{i}\hbar\, \dot{a} = [H,a]$. Alternatively, one may imagine that they arise as an approximation to the collective motion of an interacting many-particle system; as a particular example we mention density oscillations of a fluid. In the latter case, equations (\ref{eq:RPA}) go under the name of random phase approximation (RPA).

The characteristic frequencies of the dynamical system (\ref{eq:RPA}) or equivalently, the single-boson energies of the Hamiltonian $H$, can be computed as the eigenvalues of $X$. Owing to the symplectic condition $X = - J X^\mathrm{t} J^{-1}$ the characteristic polynomial satisfies $\mathrm{Det}(\lambda - X) = \mathrm{Det} (\lambda + X)$. The eigenvalues of $X$ therefore come as pairs $\pm \lambda$.

If $X$ lies at some random position in the real symplectic Lie algebra $\mathfrak{sp}_\mathbb{R}\,$, then its eigenvalues will typically be complex, since $X \in \mathfrak{sp}_\mathbb{R}$ is neither Hermitian nor anti-Hermitian. In the present context, however, complex eigenvalues are forbidden, as they correspond to the unphysical situation of unstable motion. In fact, the physical requirement of stability of the RPA dynamics dictates that the spectrum of $X$ must lie on the imaginary axis, so that the normal modes of the bosonic system are vectors in $W$ with periodic time dependence ($\propto \mathrm{e}^{-\mathrm{i} \omega t}$). Moreover, the second-quantized Hamiltonian $H$ in (\ref{eq:2ndQ}) must have a ground state in Fock space. By standard considerations of linear algebra, all these stability conditions are met if and only if $X$ lies in the set
\begin{equation}\label{eq:cone}
    \mathcal{E} := \{X\in\mathfrak{sp} \mid \mathrm{i}\Sigma_3 X > 0\}.
\end{equation}
We refer to $\mathcal{E}$ as the \emph{positive cone of elliptic generators} in $\mathfrak{sp}_\mathbb{R}\,$. It is a fact that every $X \in \mathcal{E}$ can be brought to diagonal form by a real Bogoliubov transformation, i.e.\ an element $g$ of the real symplectic group
$\mathrm{Sp}_\mathbb{R}\,$, which is defined by the condition
\begin{displaymath}
    J(g^{-1})^\mathrm{t}J^{-1}= g= \Sigma_3 (g^{-1})^\dagger \Sigma_3\,.
\end{displaymath}

\subsection{The model}\label{sect:2.2}

In the present paper we consider RPA generators $X$, or equivalently Hamiltonians $H$, which are a sum of two parts:
\begin{equation}\label{eq:mz3}
    X = K + R \,.
\end{equation}
The term $K$ is the deterministic (i.e., non-random) part of $X$. While the formalism developed below can in principle handle any choice of $K$, the explicit calculations presented in Section \ref{sect:apply} will be carried out for a simple concrete model of $K$ with unit mass matrix and elastic constants given by a discrete Laplacian. A precise description of the concrete model for $K$ is as follows.

\subsubsection{Deterministic part}\label{sect:2.2.1}

Let $\Lambda = \mathbb{Z}^d$ be a cubic lattice in $d$ space dimensions and associate with each site $j \in \Lambda$ the operators $a_j^\dagger$ and $a_j^{\vphantom{\dagger}}$ for boson creation and annihilation, respectively. We then take the second-quantized Hamiltonian to be
\begin{equation}
    H = \hbar\nu \sum_{j \in \Lambda} a_j^\dagger \, a_j^{\vphantom{\dagger}} - \frac{\hbar\nu}{4d} \sum_{\langle j, j^\prime \rangle} (a_j^{\vphantom{\dagger}} + a_j^\dagger) (a_{j^\prime}^{\vphantom{\dagger}} + a_{j^\prime}^\dagger)
\end{equation}
where the sum for the second term on the right-hand side is over nearest neighbor pairs of sites $j,j^\prime$ of $\Lambda$. Such a Hamiltonian is easily diagonalized by Fourier transforming to momentum space. The spectrum of single-boson energies $\varepsilon(k)$ as a function of the wave vector $k = (k_1,\ldots, k_d)$ comes out to be
\begin{equation}
    \varepsilon(k) = \hbar\nu \sqrt{1 - \Delta_k} \,, \qquad
    \Delta_k = \frac{1}{d} \sum_{i=1}^d \cos(k_i) .
\end{equation}
Note that $\varepsilon(k) \simeq \hbar\nu |k| / \sqrt{2d}$ for small $|k| = \sqrt{k_1^2 + \ldots + k_d^2}\,$, which tells us that the speed of sound in units of the lattice spacing is $\nu / \sqrt{2d}$.

By computing the RPA generator from the dynamical equation $\mathrm{i}\hbar\, \dot{\alpha} = [H , \alpha]$ for $\alpha = a_j^\dagger$ and $\alpha = a_j$ we obtain the expression
\begin{equation}\label{eq:mz6}
    K_1 := - \frac{\mathrm{i}\nu}{2} \begin{pmatrix} 2 -
    \Delta &-\Delta\cr \Delta & -2 + \Delta \end{pmatrix} ,
\end{equation}
where $\Delta$ is the scaled lattice Laplacian (with diagonal part removed) which has eigenvalue spectrum $\Delta_k\,$. Next, we tensor up the model by introducing $N$ identical bands. Mathematically speaking, we pass from the symplectic vector space (for each $j \in \Lambda$)
\begin{displaymath}
    \mathrm{span}_\mathbb{C} \{ a_j^{\vphantom{\dagger}} \,, a_j^\dagger\} \simeq \mathbb{C}^2
\end{displaymath}
to the tensor product $W_j := \mathbb{C}^2 \otimes \mathbb{C}^N \simeq \mathbb{C}^{2N}$ and take the generator $K$ to be
\begin{equation}
    K := K_1 \otimes \mathrm{Id}_{\mathbb{C}^N} \,.
\end{equation}
This means that creation operators $a_{j,\,n}^\dagger$ and annihilation operators $a_{j,\,n}$ get an extra band index $n = 1, \ldots, N$. Note that in the physical setting of lattice vibrations a reasonable choice of $N$ in $d$ dimensions is $N = d$ due to the vector nature of lattice displacements.

\subsubsection{Random part}

We turn to $R$, the second term in (\ref{eq:mz3}), which is random. A particular feature of our disordered model is that we take $R$ to be diagonal in the sites $j \in \Lambda$ of the lattice. For simplicity we begin the discussion with the very special case of $\Lambda$ consisting of just a single site. The full model to be discussed later is obtained by repeating the single-site discussion at every site of $\Lambda = \mathbb{Z}^d$.

With the single site of the lattice we associate a Hermitian vector space $W = \mathbb{C}^{2N}$ with symplectic structure $J = \begin{pmatrix} 0 &1_N \cr -1_N &0 \end{pmatrix}$. In order to implement the positivity condition [see Eq.\ (\ref{eq:cone})] for $X$ to be in the cone $\mathcal{E}$, we let
\begin{equation}
    R = - \mathrm{i} \Sigma_3 L^\dagger L \,, \qquad
    \Sigma_3 = \begin{pmatrix} 1_N &0 \cr 0 &-1_N \end{pmatrix} ,
\end{equation}
where $L$ is a rectangular linear operator
\begin{equation}
    L : \; W \to V \,, \qquad V = \mathbb{C}^M ,
\end{equation}
mapping $W$ into an auxiliary vector space $V$. The dimension $M$ is a parameter of our model. It may be bigger or smaller than $2N$. A special role is played by the choice $M = 2N$, as this is the minimal dimension for the operator $R$ to have full rank.

It is easy to see that for $R = - \mathrm{i}\Sigma_3 L^\dagger L$ the symplectic condition $R = - J R^{\, \mathrm{t}} J^{-1}$ holds if and only if $L$ satisfies the reality condition
\begin{equation}
    \bar{L} = L\, \Sigma_1 \;, \qquad \Sigma_1 = \begin{pmatrix} 0 &1_N\cr 1_N &0 \end{pmatrix} .
\end{equation}
This condition fixes a real form, say $U_\mathbb{R}$, of the complex vector space $U \equiv \mathrm{Hom}(W,V)$. Note that if $L^\dagger L$ has full rank then $\mathrm{i}\Sigma_3 R = L^\dagger L > 0$ and $R \in \mathcal{E}$. 

Disorder is introduced by declaring the matrix elements of $L$ to be Gaussian random variables. More precisely, we define the probability measure for $L \in U_\mathbb{R}$ as
\begin{equation}\label{eq:meas}
    d\mu(L) = C\,\mathrm{e}^{-\frac{N}{b} \mathrm{Tr}\,L^\dagger L} dL\,,
\end{equation}
where $dL$ is Lebesgue measure on the normed vector space $U_\mathbb{R}$ and $C$ is a normalization constant. The parameter $b$ is a measure of the disorder strength. We mention in passing that the model for $X = R = - \mathrm{i}\Sigma_3 L^\dagger L$ with probability measure (\ref{eq:meas}) (and $M \geq 2N$) is equivalent to the random matrix model studied in \cite{LSZ06} by different methods.

Finally, we describe the generalization to an arbitrary lattice or graph $\Lambda$. With each lattice site $j \in \Lambda$ we associate one copy $W_j$ of the Hermitian symplectic vector space $\mathbb{C}^{2N}$. The total physical space then is the orthogonal sum $W = \oplus_{j\in\Lambda} \, W_j\,$. Note that $\mathrm{dim}(W) = 2N |\Lambda|$ where $|\Lambda|$ denotes the number of sites of $\Lambda$. The full generator of the dynamics is $X = K + R$ where the deterministic part $K$ may in principle be any element of the positive cone $\mathcal{E}(W)$. For concreteness we let $\Lambda = \mathbb{Z}^d$ and take $K$ to be the generator described in Section \ref{sect:2.2.1}. The random part $R$ is a sum $R = \sum_j\, R_j$ of $R_j = - \mathrm{i}\Sigma_3 L_j^\dagger L_j^{\vphantom{\dagger}}$ made from independent and identically distributed random operators $L_j\,$. In other words, the distribution for $R$ is given by the product distribution
\begin{equation}\label{eq:meas-Lambda}
    d\mu_\Lambda (L) = \prod_{j \in \Lambda} d\mu(L_j) .
\end{equation}

\subsection{Statement of result}\label{sect:2.3}

While our interest will ultimately be in correlation functions and transport properties, we here take a first step by studying the average resolvent of the time-evolution generator $X:$
\begin{equation}\label{eq:resolvent}
    g(z) = (2N |\Lambda|)^{-1} \mathbb{E}\left( \mathrm{Tr}\,
    (z - X)^{-1} \right) ,
\end{equation}
where the symbol $\mathbb{E}(\ldots)$ means the expectation value with respect to the probability measure (\ref{eq:meas-Lambda}). Notice that by the symplectic property $X= - J X^\mathrm{t} J^{-1}$ the resolvent satisfies $\mathrm{Tr} \, (z - X)^{-1} = \mathrm{Tr}\, (z + X)^{-1}$, so $g(z) = - g(-z)$ is an odd function of the frequency parameter $z\,$. Because the support of our probability measure is contained in the positive cone of elliptic elements, $\mathcal{E}$, the eigenvalue spectrum of the random operator $X$ is always imaginary and $g(z)$ is analytic in the right and left halves of the complex $z$-plane. In the following we assume $\mathfrak{Re}\, z > 0$. It is a standard fact that the local density function $\rho$ of the characteristic boson frequencies $\omega$ can be computed from
\begin{displaymath}
    \rho(\omega) = \pi^{-1} \lim_{\epsilon\to 0+}
    \mathfrak{Re}\, g(\pm \mathrm{i}\omega + \epsilon) .
\end{displaymath}

We now come to our main result. Fixing the ratio
\begin{equation}
    a := M / 2N \,,
\end{equation}
we take the large-$N$ limit of the model with dynamical generator $X = K + R$ on $\Lambda = \mathbb{Z}^d$ as described above. We then claim that in this limit $g(z)$ is expressed by
\begin{equation}\label{eq:g(z)}
     g(z) = \frac{z}{(2\pi)^d} \int\limits_{[0,2\pi]^d} \frac{d^d k}
     {z^2 + p^2 + p\nu(2-\Delta_k) + \nu^2 (1-\Delta_k)} \,,
\end{equation}
where the complex and energy-dependent quantity $p$ is a solution of the self-consistency equation
\begin{equation}\label{eq:CPA}
     \frac{1}{b} = \frac{a}{p} - \int\limits_{[0, 2\pi]^d} \frac{d^d k}{(2\pi)^d}\; \frac{p + \nu (1 - \frac{1}{2}\Delta_k)}
     {z^2 + p^2 + p\nu(2-\Delta_k) + \nu^2 (1-\Delta_k)} \,.
\end{equation}
$p$ plays the role of a `self energy' or `coherent potential'. 
%

We briefly discuss some features of the solution in two extreme cases. There is only one relevant parameter, $b/\nu$. In the limit of weak disorder ($b \to 0$) one infers that $p \to 0$ and
\begin{equation}\label{eq:mz-2.21}
    g(z) = \frac{z}{(2\pi)^d} \int\limits_{[0,2\pi]^d} \frac{d^d k}
     {z^2 + \nu^2 (1-\Delta_k)}
\end{equation}
is simply the Cauchy transform of the local density of frequencies of the deterministic generator $K$. On the other hand, for strong disorder ($b\to\infty$) the coherent potential $p \sim b$ becomes large and the system (\ref{eq:g(z)}, \ref{eq:CPA}) simplifies to
\begin{equation}\label{eq:RMT-limit}
    g(z) = \frac{z}{z^2 + p^2} \;, \qquad
     \frac{1}{b} = \frac{a}{p} - \frac{p}{z^2 + p^2} \,.
\end{equation}
A special situation arises for $a = 1$. In this case it follows by a short computation from (\ref{eq:RMT-limit}) that the scaled function $\tilde{g} (x) := \mathrm{i}b \, g(\mathrm{i} b x)$ satisfies an equation,
\begin{displaymath}
    x = \frac{- 1}{\tilde{g} (\tilde{g}^2 - 1)} \,,
\end{displaymath}
which was derived and solved by Lueck, Sommers, and one of the authors \cite{LSZ06}. The analysis of \cite{LSZ06} shows that the density of states behaves as $x^{-1/3}$ near $x = 0$ in this case. Numerical results for $a \not = 1$ and for the more general situation of intermediate disorder strengths will be presented in Sections \ref{sect:RMT-limit} and \ref{sect:apply}.

\section{Solving the model in zero dimension}\label{sect:3}
\setcounter{equation}{0}

In the sequel we explain how to arrive at our main equations (\ref{eq:g(z)}) and (\ref{eq:CPA}). For pedagogical reasons we describe the method first for the simple situation of a single site (the zero-dimensional case). Throughout this section we let $W = \mathbb{C}^{2N}$ and $V = \mathbb{C}^M$.

\subsection{Resolvent as Gaussian integral}

Our plan is to compute the average trace of resolvent (\ref{eq:resolvent}) by a variant of the Efetov-Wegner supersymmetry method. The first step of this method is to express $\mathrm{Tr}\, (z-X)^{-1}$ as a Gaussian Berezin (super-)integral. To get started, we use the elementary identity
\begin{equation}\label{eq:mz-3.1}
    \mathrm{Tr}\,(z-X)^{-1} = \frac{\partial}{\partial z_1}
    \bigg\vert_{ z_1 = z_0 = z}\;\frac{\mathrm{Det}\, (z_1-X)} {\mathrm{Det}\,(z_0-X)}\;,
   \end{equation}
and then write each of the two determinants as a Gaussian integral -- using ordinary integration variables for the determinant in the denominator and anti-commuting variables for that in the numerator.

In the case of the ordinary Gaussian integral, there exists a convergence issue because the elements $X \in \mathfrak{sp}_\mathbb{R}$ have indefinite real and imaginary parts in general. It is therefore crucial that all our generators $X$, constrained to lie in the positive cone $\mathcal{E}$, satisfy the inequality $\mathrm{i} \Sigma_3 X > 0$. To take advantage of this positivity property, we express the determinant as follows:
\begin{align}
    \mathrm{Det} \left(z - X \right) &= \mathrm{Det}\big(z - K + \mathrm{i} \Sigma_3 L^\dagger L) = z^{2N-M} \Delta(z),\\ \Delta(z) &= \mathrm{Det} \begin{pmatrix} z\,\mathrm{Id}_V &\mathrm{i}L \cr \mathrm{i}  L^\dagger &\mathrm{i}\Sigma_3 (z^{-1} K - 1)\end{pmatrix} . \label{eq:3.3}
\end{align}
Note that owing to $\mathfrak{Re}\, z > 0$ and $\mathrm{i}\Sigma_3 K > 0$, the $2\times 2$ matrix of operators in (\ref{eq:3.3}) has positive real part.

Next, we introduce symmetric complex bilinear inner products $(\; , \;)$ for each of the two vector spaces $V$ and $W$. These inner products are consistent with the Hermitian structures of $V$ and $W$ in the sense that, e.g.\ for $V$, the sesqui-bilinear form $(v,v^\prime) \mapsto (\bar{v}, v^\prime)$ agrees with the Hermitian scalar product of $V$. We then express the reciprocal determinant $1/\Delta(z)$ as an integral over two complex vectors $v \in V$ and $w \in W:$
\begin{align}\label{eq:boson-det}
    \Delta(z)^{-1} = \int \mathrm{e}^{-z\, (\bar{v},\,v) + \mathrm{i} z^{-1} (\Sigma_3\bar{w},\,w z - K w) - \mathrm{i} (\bar{v},\, L w) - \mathrm{i} (\bar{w},\,L^\dagger v)} ,
\end{align}
where it is understood that we are integrating with the product of Lebesgue measures for $V$ and $W$. The normalization is chosen in such a way that $\int \mathrm{e}^{- (\bar{v},\,v) - (\bar{w},\,w)} = 1$. We emphasize that the integral (\ref{eq:boson-det}) converges absolutely due to $\mathrm{Re}\, z > 0$ and $K \in \mathcal{E}$.

In the case of the determinant itself we integrate in the sense of Berezin (i.e., we actually differentiate) with respect to two independent vectors $\beta$ and $\gamma$ whose components are Grassmann variables:
\begin{equation}\label{eq:fermion-det}
    \Delta(z) = \int \mathrm{e}^{z\, (\bar{\beta}, \,\beta) - \mathrm{i} z^{-1} (\Sigma_3 \bar{\gamma},\, \gamma z - K \gamma) + \mathrm{i} (\bar{\beta},\,L\gamma)+\mathrm{i}(\bar{\gamma},\,L^\dagger \beta)}
    , \qquad z \not= 0 .
\end{equation}
Again, it is understood that we are integrating with the flat Berezin form, i.e., the product of all partial derivatives w.r.t.\ the Grassmann variables. The bar in the present instance means nothing but independence, e.g., of $\bar\beta$ from $\beta$.

We now multiply the two Gaussian integral formulas (\ref{eq:boson-det}) and (\ref{eq:fermion-det}) and take the disorder average inside the absolutely convergent integral to obtain
\begin{align}
    \mathbb{E}\left( \frac{\Delta(z_1)}{\Delta(z_0)} \right) = \int
    &\mathrm{e}^{- z_0 (\bar{v}, \, v) + \mathrm{i} (\Sigma_3 \bar{w},\, w - z_0^{-1} K w) + z_1 (\bar{\beta},\, \beta) - \mathrm{i} (\Sigma_3 \bar{\gamma},\, \gamma - z_1^{-1} K \gamma)} \cr &\times \mathbb{E} \left( \mathrm{e}^{-\mathrm{i}(\bar{v},\, L w) - \mathrm{i} (\bar{w},\,L^\dagger v) + \mathrm{i}(\bar{\beta},\, L \gamma) + \mathrm{i} (\bar{\gamma},\,L^\dagger \beta)} \right) . \label{eq:mz-3.6}
\end{align}
This integral representation is a suitable starting point for further analysis.

\subsection{Taking the disorder average}

Next, we compute the disorder expectation value in (\ref{eq:mz-3.6}).
For that we introduce the quadratic quantities
\begin{align*}
    &Q := v\, (\bar{w}, \cdot) + \bar{v}\, (\Sigma_1 w, \cdot) +
    \beta (\bar\gamma , \cdot) - \bar\beta (\Sigma_1 \gamma , \cdot) ,
    \\ &Q^\ast := w\, (\bar{v}, \cdot) + \Sigma_1 \bar{w} \, (v,\cdot)
    + \gamma\,(\bar\beta,\cdot) - \Sigma_1 \bar\gamma\,(\beta ,\cdot) ,
\end{align*}
where $Q$ is meant as a linear transformation from $W$ to $V$ with coefficients in the even part of a Grassmann algebra, and similar for $Q^\ast$ with the roles of $W$ and $V$ reversed. We then have
\begin{align*}
    \mathbb{E} \left(
    \mathrm{e}^{-\mathrm{i}(\bar{v},\, L w) - \mathrm{i} (\bar{w},\, L^\dagger v) + \mathrm{i}(\bar{\beta},\, L \gamma) + \mathrm{i} (\bar{\gamma},\,L^\dagger \beta)} \right) = \int \mathrm{e}^{- \frac{\mathrm{i}}{2} \mathrm{Tr}\, (L^\dagger Q + Q^\ast L)}
    d\mu(L) = \mathrm{e}^{- \frac{b}{4N} \mathrm{Tr}\, Q^\ast Q} ,
\end{align*}
by completing the square and shifting variables.

At this point we make the observation that $\mathrm{Tr}\, Q^\ast Q$ depends on $v$, $\bar{v}$, $\beta$, $\bar\beta$ only through scalar products such as $(\bar{v},v)$, $(v,v)$, $(\bar{v},\beta)$. These share the feature of invariance under the group $\mathrm{O}_M$ of real orthogonal transformations of $V = \mathbb{C}^M$. It will be useful to organize all these $\mathrm{O}_M$-scalars into a supermatrix:
\begin{equation}\label{eq:def-P}
    P = \begin{pmatrix}
    (\bar{v} , v) &(\bar{v} , \bar{v}) &(\bar{v} , \beta)
    &- (\bar{v}, \bar\beta)\\ (v , v) &(v , \bar{v}) &(v , \beta)
    &- (v , \bar\beta)\\ (\bar\beta , v) &(\bar\beta , \bar{v})
    &(\bar\beta , \beta) &0 \\ (\beta , v) &(\beta , \bar{v}) &0
    &-(\beta , \bar\beta) \end{pmatrix} \equiv \begin{pmatrix}
    P_{00} &P_{01} \cr P_{10} &P_{11} \end{pmatrix} \;.
\end{equation}
Two of the matrix entries vanish since $(\beta,\beta) = - (\beta,\beta) = 0$ and, similarly, $(\bar\beta,\bar\beta) = 0$. We also have $(\bar\beta ,\beta) = - (\beta,\bar\beta)$. We further note the expression
\begin{displaymath}
    \mathrm{STr}\; P \equiv \mathrm{Tr}\, P_{00} - \mathrm{Tr}\, P_{11} = 2 (\bar{v},v) - 2 (\bar\beta,\beta)
\end{displaymath}
for the supertrace of $P$.

There exist certain linear dependencies amongst the matrix elements of $P$. To describe them we need the operation $P \mapsto P^{\,\mathrm{st}}$ of taking the supertranspose:
\begin{displaymath}
    \begin{pmatrix} P_{00} &P_{01}\cr P_{10} &P_{11} \end{pmatrix}^\mathrm{st} =
    \begin{pmatrix} P_{00}^{\,\mathrm{t}} &P_{10}^{\,\mathrm{t}} \cr -P_{01}^{\,\mathrm{t}} &P_{11}^{\,\mathrm{t}} \end{pmatrix} .
\end{displaymath}
With its help we can formulate the symmetries of $P$ as follows:
\begin{displaymath}
    P = \sigma P^{\,\mathrm{st}} \sigma^{-1} , \quad \sigma = \mathrm{diag}(\sigma_1,\mathrm{i}\sigma_2) , \quad
    \sigma_1 = \begin{pmatrix} 0 &1\cr 1 &0\end{pmatrix} , \quad
    \sigma_2 = \begin{pmatrix} 0 &-\mathrm{i}\cr \mathrm{i} &0 \end{pmatrix} .
\end{displaymath}

We now arrange the remaining integration variables $w$, $\bar{w}$, $\gamma$, and $\bar\gamma$ in the form of rectangular supermatrices:
\begin{displaymath}
    \widetilde{\Psi} = \left(w, \Sigma_1 \bar{w}, \gamma, - \Sigma_1 \bar{\gamma} \right) \;, \qquad \Psi = \begin{pmatrix} (\bar{w},\cdot) \cr (\Sigma_1 w,\cdot) \cr (\bar\gamma,\cdot) \cr (\Sigma_1 \gamma,\cdot) \end{pmatrix} .
\end{displaymath}
More precisely, $\Psi$ is to be viewed as a linear mapping from $W$ into the superspace $\mathbb{C}^{2|2}$ with Grassmann-even resp.\ Grassmann-odd matrix coefficients on the even resp.\ odd positions of $\mathrm{Hom} (W,\mathbb{C}^{2|2})$.
It is easy to check the identity
\begin{displaymath}
    \mathrm{Tr}\; Q^\ast Q = \mathrm{Tr}\;\widetilde{\Psi}  P\,\Psi,
\end{displaymath}
which lets us re-express our disorder average as
\begin{equation}\label{eq:mz-3.10}
    \mathbb{E} \left(\mathrm{e}^{-\mathrm{i}(\bar{v},\, Lw) -\mathrm{i} (\bar{w},\, L^\dagger v) + \mathrm{i}(\bar{\beta},\, L \gamma) + \mathrm{i} (\bar{\gamma},\,L^\dagger \beta)} \right) = \mathrm{e}^{
    -\frac{b}{4N} \mathrm{Tr}\; \widetilde{\Psi} P \, \Psi} .
\end{equation}

\subsection{Eliminating $\Psi, \widetilde{\Psi}$}

The next step is to carry out the integral over $w$, $\bar{w}$, $\gamma$, and $\bar\gamma$, thereby eliminating $\Psi$ and $\widetilde{\Psi}$ from the calculation. This will be straightforward to do because the dependence on these variables is Gaussian.

As a preparatory step, we verify from $K = - J K^\mathrm{t} J^{-1}$ the relation
\begin{equation}\label{eq:mz-3.11}
    \mathrm{i} (\Sigma_3 \bar{w}, w-z_0^{-1} Kw) - \mathrm{i}(\Sigma_3 \bar\gamma, \gamma-z_1^{-1} K \gamma) = {\textstyle{\frac{1}{2}}} \mathrm{Tr}\,( \mathrm{i} \Sigma_3 \widetilde{\Psi} \tau_3 \Psi - \mathrm{i}\Sigma_3 K \widetilde{\Psi} \hat{z}^{-1} \Psi ) ,
\end{equation}
where
\begin{equation}
    \hat{z} = \mathrm{diag}(z_0,z_0,z_1,z_1) \;, \quad
    \tau_3 = \mathrm{diag}(1,-1,1,-1) \;.
\end{equation}
By using the results (\ref{eq:mz-3.10},\,\ref{eq:mz-3.11}) in equation (\ref{eq:mz-3.6}) we then arrive at our next formula:
\begin{align*}
    \mathbb{E}\left( \frac{\Delta(z_1)}{\Delta(z_0)} \right) = \int \mathrm{e}^{- \frac{1}{2} \mathrm{STr}\,(\hat{z}\, P) + \frac{1}{2} \mathrm{Tr}\, (\mathrm{i}\Sigma_3 \widetilde{\Psi} \tau_3 \Psi - \mathrm{i}\Sigma_3 K \widetilde{\Psi} \hat{z}^{-1} \Psi - \frac{b}{2N} \widetilde{\Psi}  P\, \Psi) } .
\end{align*}
The integral on the right-hand side is still over the original variables $v$, $\bar{v}$, $\beta$, $\bar\beta$ in $P$ and $w$, $\bar{w}$, $\gamma$, $\bar\gamma$ in $\Psi$, $\widetilde{\Psi}$. Finally, by using a standard formula for Gaussian Berezin superintegrals we perform the integral over $\Psi$ and $\widetilde{\Psi}$. This results in
\begin{displaymath}
    \mathbb{E}\left( \frac{\Delta(z_1)}{\Delta(z_0)} \right) =         \int \mathrm{e}^{- \frac{1}{2} \mathrm{STr}\, (\hat{z}\, P)}
    \,\mathrm{SDet}^{-1/2}(-\mathrm{i}\Sigma_3 \otimes \tau_3 + \mathrm{i}\Sigma_3 K \otimes \hat{z}^{-1} + \mathrm{Id}_W
    \otimes b P / 2N) .
\end{displaymath}
The superdeterminant here is over the tensor product space $W \otimes \mathbb{C}^{2|2}$. We recall that the superdeterminant of a supermatrix is defined by
\begin{displaymath}
    \mathrm{SDet} \begin{pmatrix} A &B\cr C &D\end{pmatrix}
    = \frac{\mathrm{Det}(A)}{\mathrm{Det}(D - C A^{-1} B)}
    = \frac{\mathrm{Det}(A - B D^{-1} C)}{\mathrm{Det}(D)} .
\end{displaymath}

The integral above is still over the variables $v$, $\bar{v}$, $\beta$, and $\bar\beta$ entering via their scalar products into the supermatrix $P$. By scaling these integration variables so that $P \to 2N P / b$, we obtain the following expression for the generating function of our problem:
\begin{align}\label{eq:Om(z)}
    \Omega(\hat{z}) &:= \mathbb{E} \left( \frac{\mathrm{Det}(z_1-X)}
    {\mathrm{Det}(z_0-X)} \right) = (z_1 / z_0)^{2N-M}\, \mathbb{E}
    \left( \frac{\Delta(z_1)}{\Delta(z_0)} \right) = (z_0/z_1)^M \times\cr  &\times \int \mathrm{e}^{- \frac{N}{b} \mathrm{STr}\, (\hat{z}\, P)}\, \mathrm{SDet}^{-1/2}(-\mathrm{i}\Sigma_3 \otimes \hat{z} \tau_3 + \mathrm{i}\Sigma_3 K \otimes \mathrm{Id}_{2|2} + \mathrm{Id}_W
    \otimes \hat{z} P) .
\end{align}
The symbol $\mathrm{Id}_{2|2}$ stands for the identity in superspace $\mathbb{C}^{2|2}$.

\subsection{Reduction by superbosonization}

Superbosonization is a change of variables \cite{LSZ08} which lets us switch from integrating over a large number of vector-type variables, to integrating over a smaller number of matrix-type variables. In the present context these are the components of the vector variables $v$, $\bar{v}$, $\beta$, $\bar\beta$ and the matrix elements of the supermatrix $P$, respectively. Such a reduction of the number of integration variables does not come for free but requires the integrand to be invariant under one of the Lie groups $\mathrm{GL}$, $\mathrm{O}$, or $\mathrm{Sp}$. There exists a version of superbosonization for each of these classical Lie symmetries. As we have seen, our integrand is expressed in terms of quadratic invariants of the orthogonal group $\mathrm{O}_M\,$. Therefore we now recall from \cite{LSZ08} the superbosonization identity for the case of $\mathrm{O}_M$-symmetry.

The $\mathrm{O}_M$-superbosonization identity reads
\begin{equation}\label{bosonize1}
    \int F \big( P(v,\bar{v},\beta,\bar\beta) \big) =
    \int D\mu(P) \; \mathrm{SDet}^{M/2}(P) \, F(P) \;,
\end{equation}
where on the left-hand side we integrate with the flat Berezin form
\begin{displaymath}
    \prod_{m = 1}^M dv_m\, d\bar{v}_m \frac{\partial^2} {\partial\beta_m \, \partial\bar{\beta}_m} \,,
\end{displaymath}
and on the right-hand side the Berezin integration form is
\begin{equation}
    D\mu(P) = DP \; \mathrm{SDet}^{1/2}(P) ,
\end{equation}
where $DP$ is still the flat Berezin form (i.e., the product of differentials for the even variables and partial derivatives for the odd variables). The domain of integration for the so-called boson-boson block $P_{00}$ [see Eq.\ (\ref{eq:def-P})] is the space of positive Hermitian $2 \times 2$ matrices $P_{00} \equiv Q$ subject to $Q = \sigma_1 Q^{\,\mathrm{t}} \sigma_1\,$. In the fermion-fermion sector, the integration domain is the space of unitary $2 \times 2$ matrices $P_{11} \equiv U$ subject to the symmetry relation $U = \sigma_2 U^\mathrm{t} \sigma_2\,$. These matrix spaces are diffeomorphic to the symmetric spaces $\mathrm{GL}_2(\mathbb{R}) / \mathrm{O}_2$ and $\mathrm{U}_2 / \mathrm{USp}_2$ respectively.

By applying the superbosonization identity (\ref{bosonize1}) to the integral representation (\ref{eq:Om(z)}), we obtain our final result for the generating function:
\begin{align}
    \Omega(\hat{z}) = \int &D\mu(P)\; \mathrm{e}^{-\frac{N}{b} \mathrm{STr}\, P} \;\mathrm{SDet}^{M/2}(P) \cr &\times \mathrm{SDet}^{-1/2} (-\mathrm{i}\Sigma_3 \otimes \hat{z}\tau_3 + \mathrm{i}\Sigma_3 K\otimes \mathrm{Id}_{2|2} + \mathrm{Id}_W \otimes P) . \label{eq:final}
\end{align}
Notice that a substitution $P \to \hat{z}^{-1} P$ was also made. By the relation $\mathrm{SDet}^{M/2}(\hat{z}^{-1} P) = (z_1/z_0)^M \mathrm{SDet}^{M/2}(P)$ this removes the multiplicative constant $(z_0/z_1)^M$ from (\ref{eq:Om(z)}).

The result (\ref{eq:final}) is exact and mathematically rigorous for $M \geq 2$. (In the present case of $\mathrm{O}_M$-symmetry the superbosonization identity fails for $M = 1$; see \cite{LSZ08}.) From it we get the average trace of resolvent by differentiating at coinciding points $z_0 = z_1$:
\begin{displaymath}
    \mathbb{E}\left( \mathrm{Tr}\,(z-X)^{-1} \right) = \frac{\partial}{\partial z_1}\bigg\vert_{z_1 = z_0 = z}
    \Omega\big( \mathrm{diag} (z_0,z_0,z_1,z_1) \big) .
\end{displaymath}

\subsection{Random-matrix limit}\label{sect:RMT-limit}

To conclude this section we consider the special limit of vanishing deterministic generator $K=0$. In that case our expression simplifies to
\begin{align}
    \Omega(\hat{z}) = \int D\mu(P)\; &\mathrm{e}^{-\frac{N}{b} \mathrm{STr} \,P} \;\mathrm{SDet}^{M/2}(P) \cr \times &\mathrm{SDet}^{-N/2}(P - \mathrm{i}\hat{z} \tau_3)\, \mathrm{SDet}^{-N/2}(P + \mathrm{i}\hat{z}\tau_3) ,
\end{align}
where all superdeterminants and supertraces are over $\mathbb{C}^{2|2}$.
Recalling the parameter $a = M / 2N$ we see that our integral is of the form
\begin{displaymath}
    \Omega(\hat{z}) = \int D\mu(P)\; \mathrm{e}^{- N\, F(P)}
\end{displaymath}
with
\begin{displaymath}
    F(P) = b^{-1} \mathrm{STr} \,P - a\, \ln \mathrm{SDet}(P) + {\textstyle{\frac{1}{2}}} \ln \mathrm{SDet}(P - \mathrm{i}\hat{z} \tau_3) + {\textstyle{\frac{1}{2}}} \ln \mathrm{SDet}(P + \mathrm{i} \hat{z}\tau_3) .
\end{displaymath}
We now investigate the random-matrix limit $N \to \infty$ with $a = M/2N$ held fixed. In this limit the integral for $\Omega(\hat{z})$ can be computed by the saddle-point or Laplace method. By the principles of supersymmetry, the leading contributions to the integral at $z_0 = z_1$ can be shown \cite{schmittner} to come from saddle points which are multiples $P = p \, \mathrm{Id}_{2|2}$ of the identity. We here omit the details of the calculation and present only the outcome. By execution of the saddle-point method we find that
\begin{equation}\label{eq:mz15}
    g(z) := \lim_{N\to\infty} (2N)^{-1} \mathbb{E} \left( \mathrm{Tr}\,(z-X)^{-1} \right) = \frac{z}{z^2 + p^2} ,
\end{equation}
where $p$ is a solution of the saddle-point equation
\begin{equation}\label{eq:mz16}
    \frac{1}{b} = \frac{a}{p} - \frac{1/2}{p-\mathrm{i}z} - \frac{1/2}{p+\mathrm{i}z} \;.
\end{equation}
In Section \ref{sect:2.3} [see Eq.\ (\ref{eq:RMT-limit})] we already remarked that for $a = 1$ this is equivalent to an equation analyzed and solved in \cite{LSZ06}. Hence in what follows we focus on $a \not= 1$.

\begin{figure}
    \begin{center}
        \epsfig{file=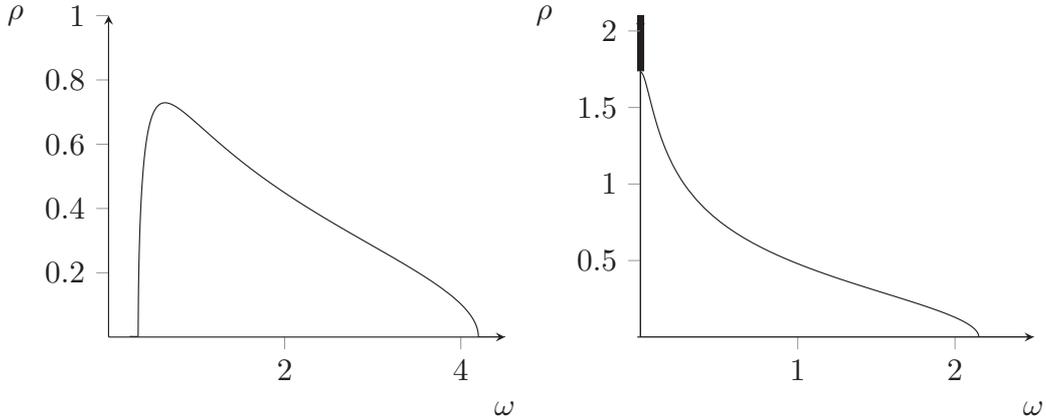,height=6cm}
    \end{center}
    \caption{Density of eigenfrequencies in the random-matrix limit $K = 0$. The parameter values are $b = 1$ and $a = 2$ (left), $a = 0.75$ (right). In the latter case there exists a Dirac-delta peak $(1-a) \delta(\omega)$ due to zero modes.} \label{fig:RMT}
\end{figure}

We begin with the case $a > 1$. A plot of the density of states for $a=2$ is shown in Figure \ref{fig:RMT}. We see that there is a gap at low frequencies.
This feature can be understood in the same way as the Marcenko-Pastur law \cite{MP-66} for rectangular Wishart matrices. Indeed, recall that our random generator is $R = - \mathrm{i}\Sigma_3 L^\dagger L$ where $L \in \mathrm{Hom}(W,V)$ is rectangular of size $M \times 2N$. The non-zero eigenvalues of $R$ coincide with those of the operator $- \mathrm{i}L\, \Sigma_3 L^\dagger$ but the latter has $M-2N = 2N (a-1)$ additional eigenvalues at zero by rank-nullity. In the large-$N$ limit the level repulsion due to this macroscopic number of zero modes produces a spectral gap of size proportional to $(a-1)$.

The gap closes as $a$ approaches unity, leading at $a = 1$ to the situation investigated in \cite{LSZ06}. For $a < 1$ it is the operator $L^\dagger L$ which by rank-nullity has $2N-M = 2N(1-a)$ zero modes, and the same goes for $R = - \mathrm{i}\Sigma_3 L^\dagger L\,$. Therefore the density of states contains a Dirac mass $(1-a)\delta(z)$ at zero in this case. A plot of the density of states for $a < 1$ is shown in Figure \ref{fig:RMT}, where see that the DOS approaches a finite value at zero frequency. (The argument of macroscopic level repulsion does not apply here, as the operator $R$ is neither Hermitian nor anti-Hermitian.)

The discussion above is concerned with the so-called bulk scaling limit. Another limit of interest is the edge-scaling limit at $z = 0$ where one sends $N \to \infty$ while keeping $z \sqrt{N}$ fixed. For $a = 1$ this limit was thoroughly investigated in \cite{LSZ06}, while for $a > 1$ the situation is trivial because of the absence of states at $z = 0$.

For $a < 1$ the edge-scaling limit was studied in \cite{schmittner}. The Hessian of the function $F(P)$ at the saddle point $P = p\, \mathrm{Id}_{ 2|2}$ has eigenvalues of order $z\,$. Therefore, in edge scaling $z \sim N^{-1/2} \to 0$ this saddle point is not isolated and one has to work with a whole supermanifold of saddle points. (Technically speaking, the saddle-point supermanifold is a Riemannian symmetric superspace $\mathrm{OSp} /\mathrm{GL}$ of type $C{\rm I}|D{\rm III}$.) The law for the density of states in the limit $N\to\infty$ turns out \cite{schmittner} to be the universal law for systems of class $D$ in the symmetry classification of \cite{AZ96}.

\section{Going beyond zero dimension}\label{sect:4}
\setcounter{equation}{0}

We now turn to the $d$-dimensional model described in Section \ref{sect:2.2}. The procedure of deriving the coherent potential approximation (\ref{eq:g(z)}, \ref{eq:CPA}) for this model remains essentially the same as before. Again, our first step is to express the determinants in (\ref{eq:mz-3.1}) as Gaussian integrals over vector variables $v$, $\bar{v}$, $\beta$, $\bar\beta$ for $V = \oplus_{j \in \Lambda} V_j$ and $w$, $\bar{w}$, $\gamma$, $\bar\gamma$ for $W = \oplus_{j \in \Lambda} W_j\,$. The Gaussian integral representation has the effect of factorizing the independent random variables associated with different sites of the lattice. The disorder average can therefore be carried out for each site separately. By the local $\mathrm{O}_M$ gauge symmetry of the model, the integrand after disorder averaging depends only on $\mathrm{O}_M$ gauge invariant combinations of the fundamental variables $v$, $\bar{v}$, $\beta$, $\bar\beta$. These organize into supermatrices $P$ as before. Thus we introduce such a supermatrix $P_j$ for each site $j \in \Lambda$ and switch to integrating over $P_j$ by the superbosonization formula (\ref{bosonize1}). Because the dependence on the variables $w$, $\bar{w}$, $\gamma$, $\bar\gamma$ is still Gaussian, they can again be integrated out to produce a superdeterminant. In this way we obtain
\begin{align}
    \Omega(\hat{z}) = \int &\prod_{j \in \Lambda} D\mu(P_j)\; \mathrm{e}^{-\frac{N}{b} \mathrm{STr}\, P_j} \; \mathrm{SDet}^{ M/2}(P_j) \cr &\times \mathrm{SDet}^{-N/2} \left( \mathrm{i}\sigma_3 K_{N=1} \otimes \mathrm{Id}_{2|2} + \sum\nolimits_j (\Pi_j \otimes P_j - \mathrm{i}\sigma_3 \otimes \hat{z}\tau_3 )\right) ,
    \label{eq:FINAL}
\end{align}
where $\Pi_j$ denotes the orthogonal projector from $W_{N=1}$ onto $(W_j)_{N=1}$. The only difference of any essence from our earlier result (\ref{eq:final}) is that the integral now is over a field of supermatrices $\{ P_j \}_{j \in \Lambda}$ instead of a single supermatrix $P$. The operator $\sum_j (\Pi_j \otimes P_j - \mathrm{i}\sigma_3 \otimes \hat{z}\tau_3)$ is diagonal on $W$ but (for generic $P_j$) non-diagonal on $\mathbb{C}^{2|2}$. On the other hand, the operator $\mathrm{i}\sigma_3 K_1 \otimes \mathrm{Id}_{2|2}$ is trivial on superspace but couples the sites of the graph $\Lambda$. The inverse square root of $\mathrm{SDet}$ is raised to the $N^\mathrm{th}$ power because each of the $N$ bands of the deterministic limit contribute the same factor.

We now face the task of analyzing the model (\ref{eq:FINAL}) by the field-theoretic methods of gradient expansion and renormalization. (Note that a closely related problem has already been tackled in \cite{schmittner}.) Hoping to make progress with this in a future publication, we here take a first step by computing the local density of states.

\subsection{DOS for the concrete model}
\label{sect:apply}

Let us finally work out the mean-field solution of the model (\ref{eq:FINAL}) with deterministic generator $K_1$ as defined in (\ref{eq:mz6}). Writing the integrand as $\mathrm{e}^{-N\, F}$ (for $M = 2N a$) we take the general variation of $F$:
\begin{align*}
    \delta F &= b^{-1} \sum\nolimits_j \mathrm{STr}\, \delta P_j - a \sum\nolimits_j \mathrm{STr}\, P_j^{-1} \delta P_j \cr &+ {\textstyle{\frac{1}{2}}} \sum\nolimits_j \mathrm{STr}\, (\Pi_j \otimes \delta P_j) \left( \mathrm{i}\sigma_3 K_1 \otimes 1 + \sum\nolimits_l (1 \otimes P_l - \mathrm{i}\sigma_3 \otimes \hat{z}\tau_3 )\right)^{-1} .
\end{align*}
For large $N$ we expect the field integral to be essentially given by a spatially homogeneous saddle point $P_j = p \, \mathrm{Id}_{2|2}$ (independent of $j \in \Lambda$) and small fluctuations around it. Therefore, after setting $z_0 = z_1 = z$ we look for solutions of $\delta F = 0$ of this very form. The variational equation $\delta F = 0$ then reduces to an equation of self-consistent mean-field type:
\begin{displaymath}
     0 = \frac{1}{b} - \frac{a}{p} + \frac{1}{2} \int \frac{d^d k} {(2\pi)^d}\, \mathrm{Tr} \begin{pmatrix} -\mathrm{i}z + p + \nu - \frac{1}{2}\nu \Delta_k &- \frac{1}{2} \nu \Delta_k \cr - \frac{1}{2}\nu \Delta_k &\mathrm{i}z + p + \nu - \frac{1}{2}\nu \Delta_k\end{pmatrix}^{-1} ,
\end{displaymath}
where we have used the property that the Laplacian $\Delta$ is diagonal with eigenvalues $\Delta_k$ in momentum space. By evaluating the trace of the matrix inverse we immediately arrive at equation (\ref{eq:CPA}). Within this mean-field (or coherent potential) approximation scheme, we obtain the expression (\ref{eq:g(z)}) for the average resolvent trace $g(z)$.

\begin{figure}
    \begin{center}
        \epsfig{file=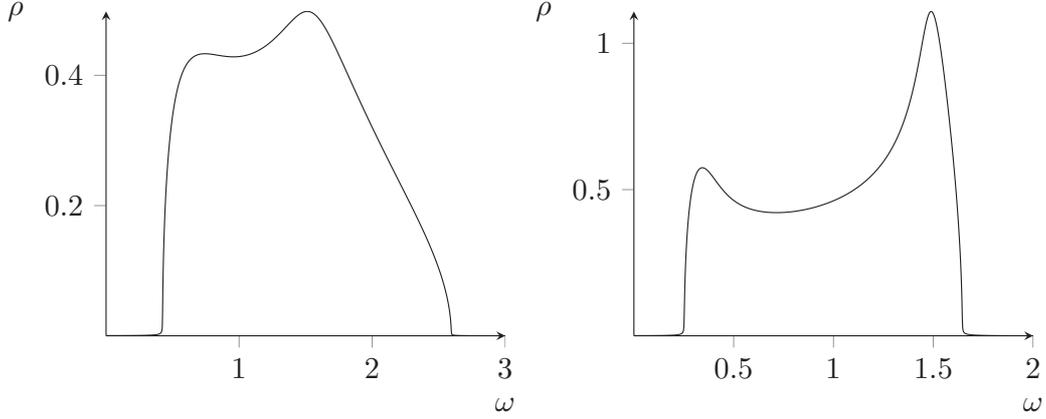,height=6cm}
    \end{center}
    \caption{DOS for $d = 1$, $\alpha = 0.75$, $\nu = 1$ and $b= 0.63$ (left), $b = 0.15$ (right).} \label{fig:CPA}
\end{figure}
Let us finish by showing some numerical results for the model in dimension $d = 1$. In this case the density of states from (\ref{eq:mz-2.21}) for the pure system ($b = 0$) is
\begin{displaymath}
    \rho(\omega) = \pi^{-1} (2\nu^2 - \omega^2)^{-1/2} .
\end{displaymath}
As is seen in Figure \ref{fig:CPA}, the van Hove singularity at $\omega = \sqrt{2} \nu$ is still visible for $b = 0.15$ (and $a = 0.75)$. As the disorder strength $b$ is increased, the bulk of the spectrum is pushed to higher frequencies and a peak begins to develop at small frequencies (see the plot for $b = 0.63$). At values of $b$ much larger than the sound velocity $\nu$ we recover the random-matrix limit shown in Figure \ref{fig:RMT}.

\smallskip\noindent\textbf{Acknowledgment.} This work was supported in part by the Deutsche Forschungsgemeinschaft (SFB/TR 12).

\end{document}